# Interstitial Iron Controlled Superconductivity in Fe$_{1+x}$Te$_{0.7}$Se$_{0.3}$

E. E. Rodriguez[1], C. Stock[1,2], P-Y Hsieh[1,3], N. Butch[4], J. Paglione[4] and M. A. Green[1,3,*]

[1] *NIST Center for Neutron Research, NIST, Gaithersburg, MD 20899 USA*
[2] *Indiana University, 2401 Milo B. Sampson Lane, Bloomington, Indiana 47408, USA*
[3] *Department of Materials Science and Engineering, University of Maryland, College Park, 20742 USA*
[4] *Center for Nanophysics and Advanced Materials, University of Maryland, College Park, MD 20742 USA*

The superconducting series, Fe(Te,Se), has a complex structural and magnetic phase diagram that is dependent on composition and occupancy of a secondary interstitial Fe site. In this letter, we show that superconductivity in Fe$_{1+x}$Te$_{0.7}$Se$_{0.3}$ can be enhanced by topotactic deintercalation of the interstitial iron, demonstrating the competing roles of the two iron sites. Neutron diffraction reveals a flattening of the Fe(Te,Se)$_4$ tetrahedron on Fe removal of iron and an increase in negative thermal expansion within the *ab* plane that correlates with increased lattice strain. Inelastic neutron scattering shows that a gapped excitation at 6 meV, evolves into gapless paramagnetic scattering with increasing iron; similar to the fluctuations observed for non-superconducting Fe$_{1+x}$Te itself.

Iron based high temperature superconductors offer new opportunities to establish the interplay between magnetism, composition and electronic properties[1]. Several structural families have now been established since the discovery of superconductivity at 26 K in LaO$_{1-x}$F$_x$FeAs[2]. These iron pnictide systems require additional cations to provide an overall Fe$^{2+}$ oxidation state, which is not necessary in the structurally simpler superconducting iron chalcogenide series, FeX (X = Te, Se, S)[3-5]. Within this series, stoichiometric FeSe is orthorhombic and superconducting at 8.5 K[3, 6], which increases to 36.7 K under high pressures[7, 8]. Fe$_{1+x}$Te can only be synthesised with large amounts of interstitial iron. This additional iron is known to greatly affect its structural and magnetic properties; at low temperature the tetragonal structure transforms to a monoclinic one with an incommensurate magnetic structure for high levels of interstitial iron (x > 0.12), compared with orthorhombic symmetry and a commensurate magnetic structure for lower iron levels[9]. Using conventional solid state chemistry methodology, it was reported that Fe$_{1+x}$Te can be formed at least over a range from Fe$_{1.076(2)}$Te to Fe$_{1.141(2)}$Te[9]. Recent attempts to deintercalate the interstitial iron proved successful with iodine as an oxidant, transforming Fe$_{1.18(5)}$Te to Fe$_{1.042(5)}$Te[10].

Previous studies on Fe$_{1+x}$Te$_{1-y}$Se$_y$ have noted that optimal superconductivity has been limited to compositions of y ~ 0.5[11]. However, a strong correlation has also been reported between anion composition and the amount of interstitial iron present for both the Fe(Te,Se)[11] and Fe(Te,S)[12] phase diagrams. The Se or S substitution reduces the lattice volume, imposes chemical pressure, and suppresses the amount of interstitial iron present. It is therefore unclear whether the particular compositions that are noted for their superconductivity are a result of optimal Te:Se ratios or whether the presence of interstitial iron, despite occupancies of only a few percent, is playing the prevailing role on the electronic properties. As for Fe$_{1+x}$Te, the effect of the interstitial iron on the structural and magnetic properties of the Fe$_{1+x}$(Te,Se) series has already been shown to be significant, particularly in the balance of the electron localization[13, 14]. Here we show that the secondary interstitial iron is the critical parameter and superconductivity in other Te:Se ratios can be induced by a secondary chemical procedure to topotactically remove the interstitial iron. Furthermore, we present inelastic neutron scattering data to demonstrate that the collective excitation at 6 meV, which has previously been reported to be present in FeTe$_{0.5}$Se$_{0.5}$ below its superconducting transition temperature[15-18], evolves to gapless paramagnetic scattering. The fact that these fluctuations are identical in Fe$_{1+x}$Te, gives further evidence that the delicate balance between superconductivity and localised magnetism is being controlled by the interstitial iron, and not the anion composition.

A powder sample of nominal composition, Fe$_{1.05}$Te$_{0.7}$Se$_{0.3}$, was synthesized by a solid state reaction of the constituent elements at 700 °C under vacuum. The energy dispersive X-ray (EDX) technique was found not to be sufficiently accurate to determine the occupancy of the iron interstitial sites. It has been reported that the occupancy values obtained for bulk single crystal X-ray diffraction determination and EDX measurements, which is a surface probe, yield different values[11]. This implies that the stability of the excess iron is less on the surface and the iron adopts more locations in the bulk of the sample. Previously we have only found single crystal X-ray diffraction or powder neutron diffraction to give reliable values of the average composition. The actual composition of the sample was determined to be Fe$_{1.048(2)}$

Te$_{0.7}$Se$_{0.3}$, by powder neutron diffraction using the BT1 diffractometer (λ = 2.0787 Å) with refinement performed using FULLPROF[19]. The composition Fe$_{1.048(2)}$Te$_{0.7}$Se$_{0.3}$ was then divided into 4 batches, three of which were exposed to different levels of I$_2$ vapour at 200°C in an evacuated glass ampoule, which topotactically deintercalates the excess iron. Powder neutron diffraction determined the compositions of these to be Fe$_{1.033(2)}$Te$_{0.7}$Se$_{0.3}$, Fe$_{1.018(2)}$Te$_{0.7}$Se$_{0.3}$ and Fe$_{1.009(3)}$Te$_{0.7}$Se$_{0.3}$. The single batch original of all samples eliminates effects from composition variation that is present in, for examples, studies that compare different Te:Se ratios.

Figure 1, panel a, shows the two iron locations in the anti-PbO structure of Fe$_{1+x}$Te$_{0.7}$Se$_{0.3}$. Magnetisation measurements were performed under zero-field cooled conditions on the four variable iron concentrations, Fe$_{1.048(2)}$Te$_{0.7}$Se$_{0.3}$, Fe$_{1.033(2)}$Te$_{0.7}$Se$_{0.3}$, Fe$_{1.018(2)}$Te$_{0.7}$Se$_{0.3}$ and Fe$_{1.009(3)}$Te$_{0.7}$Se$_{0.3}$ (Figure 1, panel b). Fe$_{1.048(2)}$Te$_{0.7}$Se$_{0.3}$ showed extremely low superconducting volume fractions. The extent of the superconducting volume fraction steadily increases with the removal of the interstitial iron, establishing a direct association between the two. The iodine deintercalation procedure is a topotactic technique that is done at very low temperatures, which are much below the synthesis temperature, thereby ruling out any possible changes to the actual Fe$_{1+x}$Te$_{0.7}$Se$_{0.3}$ framework.

Figure 2, panel a, shows the lattice parameter as a function of temperature for the four compositions, as obtained from Rietveld refinements of powder neutron diffraction data. Each of the materials undergo negative thermal expansion of the *ab* plane with the lattice parameters flat as a function of temperature below 200K, which then sharply increasing below ~125 K, whilst the c parameter shrinks in a typical linear fashion. Similar anisotropic thermal expansion has been previously reported in other members of the Fe$_{1+x}$(Te,Se) series[20, 21]. The isotropic strain of the system, obtained from the peak shape parameters of the refinements, follows identical temperature dependence. The close associated between the lattice parameter and strain suggests that the expansion in *ab*, which is directly related to the primary Fe - Fe distance, is a result of microstrain within the lattice. Analysis of the peak broadening (see supplementary information) ruled out the strain to be a result of a small orthorhombic distortion or even orthorhombic strain on a tetragonal lattice. Therefore the strain results from structural effects accumulating from the stacking of the layers and the disorder caused by the mixed Te and Se sites that have very different Fe - Te and Fe - Se bond distances or a slight distortion to a symmetry lower than orthorhombic. It is interesting to note that there is an overall increase in the lattice parameters on reduction of the interstitial iron; the composition with lower iron content are actually larger, reflecting the attractive bonding involving the interstitial iron, both within and between, the mostly van der Waals layers. This contrasts with a reduction of both the Fe - Te and Fe - Se bond distances (Figure 2, panel c) on iron deintercalation; the removal allows for the relaxation in the *ab* plane, but the reduction of bonding between the interstitial iron and the anion allows the primary tetrahedral iron to establish stronger bonding and therefore shorter bond distances to the anions. A summary of the structural changes present on iron removal are given in Figure 3, panel a.

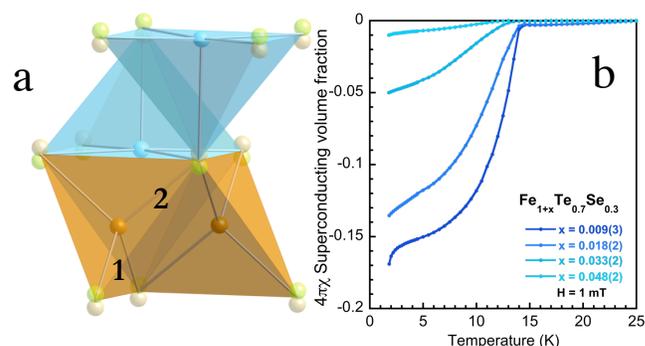

Figure 1, panel **a** Structure of Fe$_{1+x}$Te$_{0.7}$Se$_{0.3}$, highlighting the two distinct iron locations; **1** forms the edge shared tetrahedral layers, while the partially occupied (< 5%) square pyramidal site **2** lies within the (Te, Se) split site plane. Panel **b** interstitial iron in Fe$_{1+x}$Te$_{0.7}$Se$_{0.3}$ is the dominant factor controlling the balance between superconductivity and localised magnetism as measured by a commercial SQUID magnetometer.

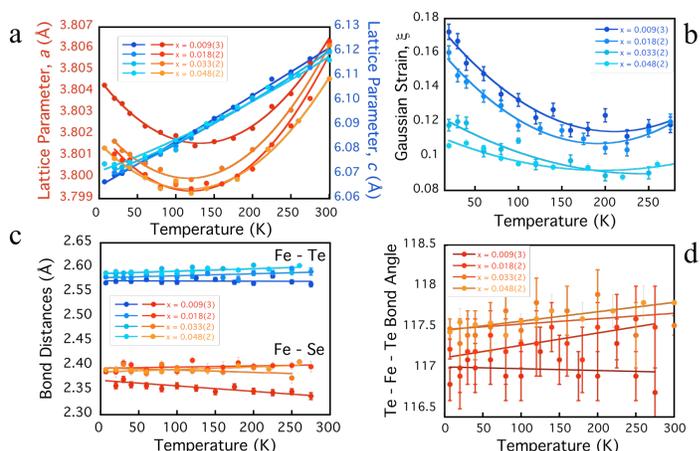

Figure 2 panel **a** Lattice parameter temperature dependence of Fe$_{1+x}$Te$_{0.7}$Se$_{0.3}$, showing the negative thermal expansion in the *ab* plane is associated with **b** gaussian strain in the lattice. **c** Both the Fe - Te and Fe - Se bonds increase on deintercalation whilst **d** the Te- Fe -Te bond angles widen resulting in a flattened tetrahedron. Error bars represent one standard deviation.

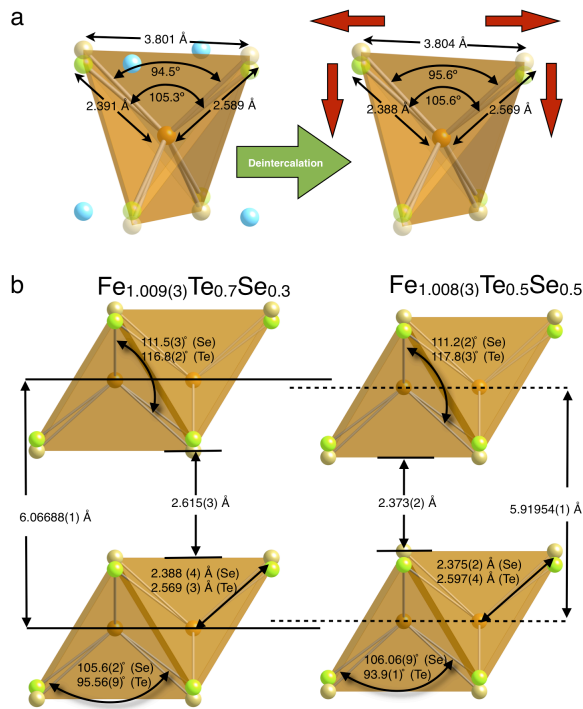

Figure 3 a Comparison of the structures of $Fe_{1+x}Te_{0.7}Se_{0.3}$ before and after removal of the interstitial iron with iodine, showing the deintercation process increases the Fe - Fe distances while squashing the Fe tetrahedron in the c direction. b Comparison of the structures of $Fe_{1.009(3)}Te_{0.7}Se_{0.3}$ with $Fe_{1.008(3)}Te_{0.5}Se_{0.5}$. The greatly increased $c$ parameter in the 70:30 over the 50:50 composition of around 0.15 Å, combined with the collapse of the $Fe(Te_{0.7}Se_{0.3})$ tetrahedron demonstrated in the less acute Te - Fe - Te bond angles and shorter Fe - Te bond distances, results is a significantly increased interlayer separation of ~ 0.24 Å. The superconducting $T_c$ is unaffected by these structural changes.

As a comparison with $Fe_{1.009(3)}Te_{0.7}Se_{0.3}$, we determined by powder neutron diffraction, the structure of $Fe_{1.008(3)}Te_{0.48(1)}Se_{0.52(2)}$ and find some stark contrasts between their structures. The former has lattice parameters of $a$ = 3.80429 (4) Å and $c$ = 6.06688 (9) Å at 5 K, whereas the latter possesses lattice parameters of $a$ = 3.79496 (5) Å and $c$ = 5.91966 (12) Å at 5 K. The Fe - Fe distance is directly related to the lattice parameter, such that $d_{Fe-Fe} = a/\sqrt{2}$. Therefore, the small change in $a$ of ~ 0.01 Å, suggests that the Fe - Fe distances within the $ab$ plane are relatively insensitive to the anion composition. In contrast, significant differences in bond distances along the c direction are observed; the 70:30 composition has significant shrinkage in the Fe tetrahedron or intralayers, but an increase between the van der Waals gap or interlayer spacing of ~ 0.24 Å, which results in an overall increase in $c$ of ~ 0.15 Å, when compared with the 50 : 50 composition. As the superconducting transition temperatures are very similar in both compositions at ~14 K, this would imply that the layer distances, and parameters related to this distance such as anion height, are not critical parameters in controlling the superconductivity. A detailed comparison of the two Fe environments is given in Figure 3, panel b.

To understand the effects of deintercalation on the magnetic fluctuations in $Fe_{1+x}$ (Te,Se), we performed inelastic neutron scattering measurements using the DCS spectrometer at NIST an incident wavelength of 2.4 Å for $Fe_{1-x}Te_{0.7}Se_{0.3}$ and 1.8 Å for $Fe_{1.06}Te$; the results are summarized in Figure 4. The sample with the lowest concentration of excess iron and largest superconducting volume fraction shows a distinct feature centered at a $|Q| \approx 1.4$ Å$^{-1}$ and with an energy gap of 6 meV (Figure 4, panel a). This excitation has previously been observed below the superconducting transition temperature in the compositions close to $FeTe_{0.5}Se_{0.5}$ and is thought to be associated with the superconducting state[1, 15-17]. As the interstitial iron concentration is increased, paramagnetic fluctuations fill in the energy gap, starting from $|Q| \approx 0.9$ Å$^{-1}$ close to the elastic line and dispersing towards the position of the gapped excitation (Figure 4, panels b to d). For the sample with the maximum amount of excess iron, $Fe_{1.048(2)}Te_{0.7}Se_{0.3}$, this paramagnetic scattering completely overwhelms any gapped excitation that may have been present. The inelastic spectra clearly indicate that small changes in the interstitial iron concentration, even by a few percent, are critical to the electronic and magnetic properties of the $Fe_{1+x}(Te,Se)$ superconductors.

There is substantial evidence pointing towards a direct coupling between antiferromagnetism and superconductivity in iron based superconductors. Mostly notably, strong magnetic collective excitations have been observed in both $Ba_{0.6}K_{0.4}Fe_2As_2$[22] and Fe(Te,Se) [15-17] systems and have been thought to directly result from the presence of an electronic superconducting energy gap. The collective mode has been found to be gapped and to draw spectral weight from lower energies in these systems, conserving sum rules required in neutron scattering. Furthermore, inelastic neutron studies of other superconductors where magnetism is thought to play an important role, such as $CeCu_2Si_2$[23], $CeCoIn_5$[24], and $YBa_2Cu_3O_{6.5}$[25], there is a clear shift in spectral weight from the gapped excitation to gapless magnetic fluctuations as these systems becomes non-superconducting. Therefore, the gapped fluctuations present in panels a) and b) in Figure 4 point towards strong evidence that superconductivity and magnetism are directly coupled in these systems. For samples which are not bulk superconductors (panels c) and d)), the gapped magnetic fluctuations are replaced by gapless paramagnetic fluctuations. Our results demonstrate a direct coupling between superconductivity and antiferromagnetism and, most importantly, the iron doping on the interstitial site. These results also demonstrate a change in the wavevector of fluctuations with doping on the interstitial site. In previous neutron scattering experiments, the gapped excitation has been found to occur at a wavevector of $\mathbf{Q} = [½, ½]$ within the *ab* plane; this ordering corresponds to antiferromagnetic coupling along a diagonal of the Fe square sublattice. However, the magnetic ordering within $Fe_{1+x}Te$ occurs with a $\mathbf{Q} = [½, 0]$, or antiferromagnetic coupling along one side of the Fe square sublattice[9]. For comparison, we include in Figure 4, panel e, the inelastic spectrum of non-superconducting $Fe_{1.06}Te$, which also shows gapless magnetic scattering. It is important to note that $Fe_{1.06}Te$ contains long-ranged magnetic ordering while the $Fe_{1+x}(Te,Se)$ samples do not. However, the gapless magnetic fluctuations observed in the samples with high amounts of interstitial iron corresponds to a |$\mathbf{Q}$| value close to that in $Fe_{1+x}Te$. While the data is powder averaged, the wavevectors are consistent with previous single crystal work, which demonstrates that the magnetic excitations change wavevector from [½ 0] to [½ ½] in the presence of superconductivity[16, 18]. Indeed, some density functional theory calculations studies have shown that the interstitial iron plays a crucial role on whether [½, ½]- or [½, 0]-type magnetic interactions will dominate[26]. Thus, our results further advance the theory that magnetic interactions along $\mathbf{Q} = [½, 0]$, promoted by the interstitial iron sites, are antagonistic to superconductivity in the $Fe_{1+x}(Se,Te)$ series.

A comparison between the two superconducting compositions of $Fe_{1.009(3)}Te_{0.7}Se_{0.3}$ and $Fe_{1.008(3)}Te_{0.48(1)}Se_{0.52(2)}$ revealed some notable differences between their two structure. Nevertheless they show similar superconducting transition temperatures and very similar excitation spectra, which demonstrates the robustness of superconductivity in the Fe(Te,Se) series. It establishes that superconductivity can occur with different compositions and geometry as long as sufficient interstitial iron that destroys superconductivity is removed; demonstrating the critical role the iron concentrations and not compositions has on the superconducting properties. It will be interesting to evaluate the full phase diagram to establish whether $Fe_{1+x}Te$ or composition close to $Fe_{1+x}Te$ can support superconductivity given sufficient removal of interstitial iron.

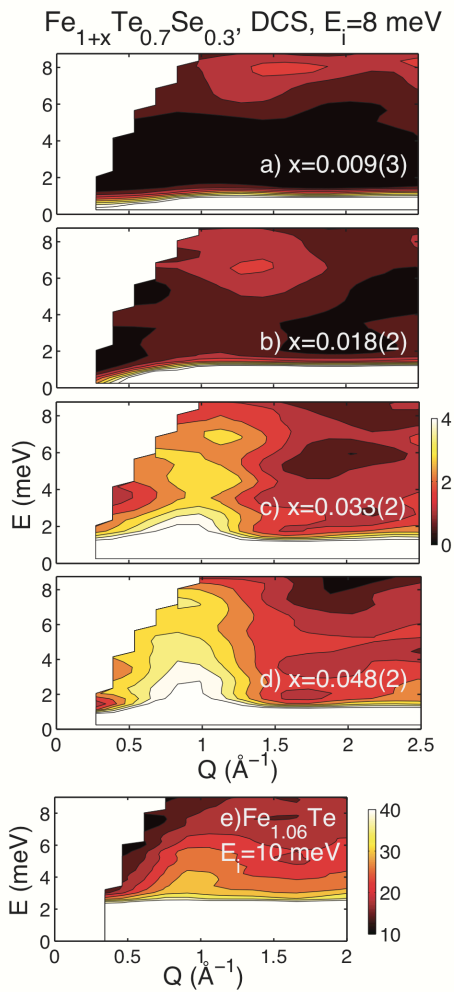

Figure 4 Panel a, Evolution of the excitation spectrum in $Fe_{1+x}Te_{0.7}Se_{0.3}$, showing the gapped excitation at 6 meV for x = 0.009(3) disappearing to gapless paramagnetic scattering on increase of interstitial iron to x = 0.048(2). b Gapless paramagnetic scattering in $Fe_{1.06}Te$ similar to excitations present in composition with large amounts of interstitial iron.